\documentstyle[osa,manuscript]{revtex}

\begin{document}
\title{Magnetic relaxation and magnetization field dependence measurements in La$%
_{0.5}$Ca$_{0.5}$MnO$_{3}$}
\author{J. L\'{o}pez*, P. N. Lisboa-Filho, W. A.\ C. Passos, W. A. Ortiz and F. M.
Araujo-Moreira}
\address{Grupo de Supercondutividade e Magnetismo,\\
Departamento de F\'{i}sica, Universidade Federal de S\~{a}o Carlos, Caixa
Postal - 676, S\~{a}o Carlos, SP, 13565-905, BRAZIL, *jlopez@df.ufscar.br}
\maketitle

\begin{abstract}
We reported a systematic change in the average magnetic relaxation rate,
after the application and removal of a 5 T magnetic field, in a
polycrystalline sample of La$_{0.5}$Ca$_{0.5}$MnO$_{3}$. Magnetic relaxation
measurements and magnetization versus field curves were taken from 10 K to
160 K. The long time behavior of the relaxation curves was approximately
logarithmic in all cases.

\end{abstract}

Charge ordering compounds have a large variation of resistivity and
magnetization as a function of temperature and magnetic field\cite{Xiao}$^{,}
$ \cite{Moritomo}. A representative example, La$_{0.5}$Ca$_{0.5}$MnO$_{3}$
has a paramagnetic-ferromagnetic transition around 265 K and a
ferromagnetic-antiferromagnetic (FM-AFM) phase transition at 160 K\cite
{Huang}. Accompanying the FM-AFM transition there is also a charge
disordered to a charge ordered phase transition\cite{Huang}.

Polycrystalline samples of La$_{0.5}$Ca$_{0.5}$MnO$_{3}$ were prepared by
the solid-state method described elsewhere\cite{Roy2}. X-ray diffraction
measurements pointed out high quality samples. Magnetization measurements
were done with a standard MPMS-5S SQUID magnetometer. The relaxation
measuring procedure was the following: first, the sample was heated to 400 K
in zero magnetic field; second, the remanent magnetic field in the solenoid
of the SQUID magnetometer was set to zero; third, the sample was cooled down
to the working temperature in zero magnetic field; fourth, an applied
magnetic field (H) was increased from 0 to 5 T at a rate of 0.83 T/minute
and remained applied for a waiting time t$_{w}$=50 s; fifth, H was decreased
to zero at the same rate; finally, when H was zero (we defined this time as
t=0) the M(t) curve was recorded for more than 210 minutes.

Figure 1 shows magnetic relaxation measurements from 10 K to 160 K. To
facilitate the comparison between curves at different temperatures, the
magnetization in each case has been normalized to the corresponding value at
t=0. As can be seen, the average relaxation rate (mean slope of each curve)
decreases systematically with increasing temperatures from 10 K to 150 K. It
is important to note that slopes here are negatives. The absolute ratio of
magnetization change, between the initial and the last measurement made, at
10 K is only 1 \%, while at 150 K is about 20 \%. This qualitative behavior
is expected because an increase in the thermal energy should produce a
faster relaxation in the magnetization. However, as was reported before\cite
{J. López-PRB}, the average relaxation rate increases between 150 K and 195
K. The curve corresponding to 160 K, that illustrates this point, is
included in figure 1.

Figure 2 shows representative measurements of the magnetization field
dependence. These curves have been explained considering two phases:
ferromagnetic droplets immersed in the charge ordered antiferromagnetic
matrix\cite{Khomskii}. All curves present a rapid increase in the
magnetization at low field values, due to the orientation of the
ferromagnetic droplets. At about 0.4 T the magnetization starts to increase
at a slower rate, showing the gradual annihilation of the antiferromagnetic
phase. A small hysteresis is found in the curves from 10 to 140 K. However,
for 150 K and 160 K, the magnetization increases rapidly for field above 3 T
and the hysteresis grows, reflecting the complete destruction of the
antiferromagnetic phase.

The relaxation measurements at long time scales follow approximately a
logarithmic law. This logarithm-like relaxation has been found in spin glass
systems\cite{Mydosh} and mixture of small ferromagnetic particles\cite
{Labarta}. The logarithmic relaxation has been attributed to the
distribution of energy barriers separating local minima, which correspond to
different equilibrium states\cite{Mydosh}$^{,}$ \cite{Labarta}. As seen in
figure 2, our sample of La$_{0.5}$Ca$_{0.5}$MnO$_{3}$ has a mixture of
ferromagnetic and antiferromagnetic domains, which produce frustration in
the interactions among individual spins. A systematic fitting of all
relaxation curves will be published elsewhere.

Concluding, we presented a systematic decrease in the average magnetic
relaxation rate from 10 K to 150 K in a polycrystalline sample of La$_{0.5}$%
Ca$_{0.5}$MnO$_{3}$. The long time behavior of the relaxation curves was
approximately logarithmic in all cases. We thank FAPESP, CAPES, CNPq and
PRONEX for financial support.

Figure 1. Normalized magnetic relaxation measurements after applying and
removing an applied magnetic field of 5 T. Time is shown in logarithmic
scale. The large arrow indicates the direction of increasing temperatures
and decreasing relaxation rates (slope more negative) between 10 K and 150
K. Also shown is the curve for 160 K that presents a higher relaxation rate
in comparison with the one at 150 K.

Figure 2. Magnetization versus applied field curves for temperatures around
150 K and one representative curve of the low temperature behavior.
Magnetization is given in Bohr magnetons per manganese ion.

\end{document}